 \documentclass[final,5p,times,twocolumn]{elsarticle}

\usepackage{amsmath} 
\usepackage{amsfonts}   
\usepackage{amssymb}
\usepackage{lipsum}
\usepackage{epsfig}                   
\usepackage{stmaryrd}   
\usepackage{esvect} 
\usepackage{comment}
\usepackage{color}
\usepackage{hyperref}

\usepackage{eso-pic}
\newcommand{\reportnum}[2]{
  \AddToShipoutPictureBG*{%
    \AtPageUpperLeft{%
      \hspace{0.75\paperwidth}%
      \raisebox{#1\baselineskip}{%
        \makebox[0pt][l]{\textnormal{#2}}
  }}}%
}

\DeclareMathOperator{\tr}{tr}		
\DeclareMathOperator{\ad}{ad}		
\DeclareMathOperator{\Lie}{Lie}		
\DeclareMathOperator{\U}{\mathit{U}}

\newcommand{\AdS}{\mathrm{AdS}}

\newcommand{\GL}[1]{\mathrm{GL}(#1)}
\newcommand{\SL}[1]{\mathrm{SL}(#1)}
\newcommand{\SU}[1]{\mathrm{SU}(#1)}
\newcommand{\Uni}[1]{\mathrm{U}(#1)}
\newcommand{\ISO}[1]{\mathrm{ISO}(#1)}
\newcommand{\SUstar}[1]{\mathrm{SU}^{*}(#1)}
\newcommand{\Spin}[1]{\mathrm{Spin}(#1)}

\newcommand{\USp}[1]{\mathrm{USp}(#1)}

\newcommand{\SO}[1]{\mathrm{SO}(#1)}

\newcommand{\Ex}[1]{\mathrm{E}_{#1}}

\newcommand{\Gx}[1]{\mathrm{G}_{#1}}

\newcommand{\miniworld}[1]{\begin{minipage}[c]{0.3\linewidth}  \vspace{0.2cm} 
\centering
#1      
\vspace{0.2cm} \end{minipage}}
\newcommand{\tinyworld}[1]{\begin{minipage}[c]{0.03\linewidth}  \vspace{0.2cm} 
\centering
#1      
\vspace{0.2cm} \end{minipage}}

\newcommand{\beq}{\begin{equation}}
\newcommand{\eeq}{\end{equation}}

\journal{Physics Letters B}

\begin{document}

\begin{frontmatter}
\reportnum{-6}{HU-EP-23/55}

\title{Vacua scan of $5d$, $\mathcal{N}=2$ consistent truncations}


\author[a]{Grégoire Josse}
\ead{gregoire.josse@physik.hu-berlin.de}
\affiliation[a]{organization={Institut für Physik, Humboldt-Universität zu Berlin, IRIS Gebäude},
            addressline={Zum Großen Windkanal 2}, 
            city={Berlin},
            postcode={12489}, 
            country={Germany}}
\author[b]{Francesco Merenda}
\ead{fmerenda@lpthe.jussieu.fr}
\affiliation[b]{organization={Laboratoire de Physiaue Théorique et Hautes Energies, Sorbonne Université},
            addressline={4 place Jussieu}, 
            city={Paris},
            postcode={75005}, 
            country={France}}

\begin{abstract}
In this letter we present a scan for new vacua within consistent truncations of eleven/ten-dimensional supergravity down to five dimensions that preserve $\mathcal{N}=2$ supersymmetry, after their complete classification in \cite{JosseMalekPetriniWaldram_2021}. We first make explicit the link between the equations of exceptional Sasaki-Einstein backgrounds in \cite{AshmorePetriniWaldram_2016} and the standard BPS equations for 5d $\mathcal{N}=2$ of \cite{LouisMuranaka_2016}. This derivation allows us to expedite a scan for vacua preserving $\mathcal{N}=2$ supersymmetry within the framework used for the classification presented in \cite{JosseMalekPetriniWaldram_2021}.
\end{abstract}


\end{frontmatter}




\section{Introduction}
\label{introduction}

When compactifying ten- or eleven-dimensional supergravity, the low-energy limits of type II string theory and M-theory, a strategy to obtain the underlying effective field theory is by integrating out modes that are above a certain energy scale. This is only viable if the theory features scale separation. If this is not the case, like in AdS compactifications, or if one needs to also keep some massive modes, another method is given by consistent truncations.

The key idea is that after the compactification we only retain a subset of fields. This is done in a consistent way, in the sense that in the lower dimensional field theory the equations of motion and BPS equations must not source any of the truncated fields.

In a previous work \cite{JosseMalekPetriniWaldram_2021}, a classification of consistent truncations of ten- or eleven-dimensional supergravity down to $\mathcal{N}=2$ gauged supergravity in five dimensions with vector, tensor and hypermultiplets was provided. In this letter we will first lay down an explicit connection of the exceptional Sasaki-Einstein equations of \cite{AshmorePetriniWaldram_2016} to the BPS equations of five-dimensional supergravity with $\mathcal{N}=2$ given in \cite{LouisMuranaka_2016}. We will then rephrase these BPS equations through the formalism of  \cite{JosseMalekPetriniWaldram_2021}. Finally, we will showcase the truncations which have been found to admit an AdS $\mathcal{N}=2$ vacuum, specifying their gauging.

This letter is structured as follows. We will start by reviewing the properties of $\mathcal{N}=2$ gauged supergravity in five dimensions. We will also briefly recall the framework of exceptional generalised geometry, at the foundation of the works of \cite{AshmorePetriniWaldram_2016} and \cite{JosseMalekPetriniWaldram_2021}. We will then review the way consistent truncations are built in general, before specifying to $\mathcal{N}=2$ truncations. We will describe how to bridge the exceptional Sasaki-Einstein equations of \cite{AshmorePetriniWaldram_2016} with the BPS equations of $\mathcal{N}=2$ five-dimensional supergravity from \cite{LouisMuranaka_2016}. Lastly, we will present the results of our vacua search.

\section{$\mathcal{N}=2$ gauged supergravity in five dimensions}

The gravity multiplet contains the graviton $e^a_\mu$, two gravitini $\psi^{x}_\mu$ transforming as a doublet of the R-symmetry group $\SU{2}_R$ and a vector field, the graviphoton $A_\mu$. We will denote the $\SU{2}_R$  indices by $x=1,2$, while the index $a$ denotes flat space-time indices and $\mu=1\ldots 5$ the curved spacetime indices.

We could also have extra matter multiplets: a vector multiplet consists of a vector $A_\mu$, an $\SU{2}_R$ doublet of spin-1/2 fermions $\lambda^x$ and a complex scalar $\phi$. In a tensor multiplet, the vector is dualised to an anti-symmetric two-form $B_{\mu\nu}$. A hypermultiplet contains four real scalars $q^u$ ($u=1,\dots,4$) and a pair of spin-1/2 fermions $\zeta^x$, also transforming as a doublet of the R-symmetry group.

Consider a theory with $n_{\rm V}$ vector multiplets, $n_{\rm T}$ tensor multiplets  and $n_{\rm H}$ hypermultiplets. The complete field content of the theory will then be
\begin{align}
\label{total_content}
\{ e^a_\mu, \psi^x_\mu , A^I_\mu, B^M_{\mu \nu}, \phi^i, q^X,  \lambda^{x i}, \zeta^{x A} \},  
\end{align}
where $A^I_\mu$, with $I=0,\dots,n_{\rm V}$, denotes the graviphoton and the  vector fields in the vector multiplets; $B^M_{\mu \nu}$, with $M= n_{\rm V} + 1, \dots, n_{\rm V} + n_{\rm T} =: n_{\rm VT}$, represent the two-forms in the tensor multiplets. 
The scalars of the vector/tensor multiplets are collectively denoted by $\phi^i$ with $i=1, \dots,  n_{\rm VT}$, while the scalars in the hypermultiplets are denoted by  $q^X$, with $X=1,\dots,4 n_{\rm H}$. Finally, the fermions in the vector/tensor and hypermultiplets are  $\lambda^{x i }$ and  $\zeta^{x A}$, where   $A=1,\dots, 2n_{\rm H}$ is a $\USp{2 n_{\rm H}}$ index.

The vector/tensor scalars $\phi^i$ parameterise a very special real manifold $\mathcal{M}_{\rm{VT}}$, while the hyper-sector scalars parameterise a quaternionic-K\"ahler manifold $\mathcal{M}_{\rm{H}}$. 
The total scalar field space $\mathcal{M}$, parameterised by $(\phi^i,q^X)$, splits into
\begin{align}
\label{total_scalar_mfld}
\mathcal{M}=\mathcal{M}_{\rm VT} \times \mathcal{M}_{\rm H}.
\end{align}
It is convenient to see $\mathcal{M}_{\rm VT}$ as an $n_{\rm VT}$-dimensional  hypersurface embedded in an $(n_{\rm VT}+1)$-dimensional ambient space spanned by the embedding coordinates $h^{\tilde{I}}=h^{\tilde{I}}(\phi^i)$, with $\tilde{I}=0,\dots,n_{\rm VT}$. The hypersurface is defined by the cubic,
\begin{align}
    \label{cubic_hypersurface}
    C_{\tilde{I}\tilde{J}\tilde{K}} h^{\tilde{I}}h^{\tilde{J}}h^{\tilde{K}}=1,
\end{align}
where $C_{\tilde{I}\tilde{J}\tilde{K}}$ is a completely symmetric, constant tensor. 

The metric on the $(n_{\rm VT}+1)$-dimensional ambient space is given by
\begin{align}
    \label{ambient_space_metric}
    a_{\tilde{I}\tilde{J}} = 3 h_{\tilde{I}}h_{\tilde{J}} - 2 C_{\tilde{I}\tilde{J}\tilde{K}} h^{\tilde{K}},
\end{align}
where 
\begin{equation}
 \label{h_shortcuts_2}
 h_{\tilde{I}} \equiv C_{\tilde{I}\tilde{J}\tilde{K}} h^{\tilde{J}}h^{\tilde{K}}=a_{\tilde{I}\tilde{K}} h^{\tilde{K}} \, .
\end{equation}
The metric $a_{\tilde{I}\tilde{J}}$ is related to the metric  on $\mathcal{M}_{\rm VT}$  by 
 the pull-back 
\begin{align}
    \label{MVT_metric}
    g_{ij} = h^{\tilde{I}}_i h^{\tilde{J}}_j a_{\tilde{I}\tilde{J}},
\end{align}
where $h^{\tilde{I}}_i$ is defined as 
\begin{align}
    \label{h_shortcuts_1}
    & h^{\tilde{I}}_i \equiv - \sqrt{\frac{3}{2}} \partial_i h^{\tilde{I}}
\end{align}

From \eqref{cubic_hypersurface} and \eqref{h_shortcuts_2}, it 
follows that 
\begin{equation}
\label{h_properties}
h_{\tilde{I}}h^{\tilde{I}}= 1, \qquad  h_{\tilde{I}} h^{\tilde{I}}_i = 0, \qquad h^{\tilde{I}} h_{\tilde{I}i} =0, \qquad h_{\tilde{I}}h^{\tilde{I}}_i=0.
\end{equation}
Combining \eqref{h_properties} and \eqref{ambient_space_metric} we can rewrite the metric $g_{ij}$ as
\begin{align}
    \label{MVT_metric_refined}
    g_{ij}= - 2 h^{\tilde{I}}_i h^{\tilde{J}}_j C_{\tilde{I}\tilde{J}\tilde{K}} h^{\tilde{K}}
\end{align}

$\mathcal{M}_{\rm H}$ is endowed with a metric $g_{XY}$ and a triplet of complex structures ${(J^\alpha)_X}^Y$, $\alpha=1,2,3$, such that
\begin{align}
    \label{MH_complex_structures}
    [J^\alpha, J^\beta]= 2 \epsilon^{\alpha \beta \gamma} J^\gamma, \qquad (J^\alpha)^2 = - \rm{Id},
\end{align}
with respect to which the metric $g_{XY}$ is hermitian.

We denote by $\mathcal{V}$ the $(n_{\rm VT}+1)$-dimensional vector space consisting of the graviphoton, the $n_{\rm V}$ vectors and the $n_{\rm T}$ tensors. The gauging defines a Leibniz algebra $\mathfrak{v}$ on $\mathcal{V}$, that is a bilinear bracket $\llbracket v,w \rrbracket$ satisfying a Leibniz relation:
\begin{align}
    \label{leibniz_relation}
    \llbracket u, \llbracket v,w \rrbracket \rrbracket =\llbracket \llbracket u,v \rrbracket, w \rrbracket + \llbracket v, \llbracket  u,w   \rrbracket \rrbracket, \qquad \forall \, u,v,w \in \mathcal{V}.
\end{align}
The algebra $\mathfrak{v}$ defines some structure constants ${t_{\tilde{J}\tilde{K}}}^{\tilde{I}}$ via
\begin{align}
    \label{leibniz_algebra_structure_csts}
    {\llbracket v,w \rrbracket}^{\tilde{I}} = {t_{\tilde{J}\tilde{K}}}^{\tilde{I}} v^{\tilde{J}} w^{\tilde{K}}, \qquad \forall \, v,w \in \mathcal{V}.
\end{align}
This bracket does not define a Lie algebra, since it is not necessarily antisymmetric: ${t_{\tilde{J}\tilde{K}}}^{\tilde{I}} \neq -{t_{\tilde{K}\tilde{J}}}^{\tilde{I}}$. As explained in more detail in \cite{JosseMalekPetriniWaldram_2021}, we can choose a particular splitting so that $\mathcal{V}=\mathcal{R}\oplus\mathcal{T}$ and fix a basis labelling components in $\mathcal{R}$ by the index $I=0,\dots,n_{\rm V}$ and components in $\mathcal{T}$ by the index $M=n_{\rm V} +1, \dots, n_{\rm VT}$: this structure will result in the relations
\begin{align}
    \label{t_relations}
    {t_{IJ}}^K = {f_{IJ}}^K, \qquad {t_{M\tilde{I}}}^{\tilde{J}}=0, \qquad {t_{(I\tilde{J})}}^I=0,
\end{align}
where ${f_{IJ}}^K=-{f_{JI}}^K$ are the structure constants of $\mathfrak{g}_{\rm{ext}}$. The split into vector and tensor multiplet is thus completely defined by the choice of Leibniz algebra. 

Given any $v \in \mathcal{V}$, the adjoint action
\begin{equation}
  \begin{aligned}
    \label{adjoint_action_V}
    t_v \, : \, \mathcal{V} &\to \mathcal{V},\\
              w & \mapsto t_v w := \llbracket v,w \rrbracket
\end{aligned}
\end{equation}
defines a Lie algebra. In terms of the split basis, the generators are \cite{BergshoeffCucu_2002}
\begin{align}
    \label{t_generators_matrix_form}
    {(t_I)_{\tilde{J}}}^{\tilde{K}}= \begin{pmatrix}
{(t_I)_J}^K & {(t_I)_J}^N \\
0 & {(t_I)_M}^N 
\end{pmatrix},
\end{align}
such that 
\begin{align}
    \label{t_generators_structure_constants}
    [t_I, t_J]=-{f_{IJ}}^K t_K.
\end{align}

The tensors $(t_I)_{\tilde J}{}^{\tilde K} $ can also be expressed in terms of the embedding tensor \cite{Samtleben_2008,Trigiante_2016}. This is a map
\begin{align}
\Theta \, : \, \mathcal{V} \to \mathfrak{g}_{\rm iso}, 
\end{align}
where $\mathfrak{g}_{\rm iso}$ is a the Lie algebra of isometries of the scalar field manifold. For $\mathcal{N}=2$ supersymmetry, the isometry algebra splits as $\mathfrak{g}_{\rm iso}=\mathfrak{g}_{\rm VT}\oplus\mathfrak{g}_{\rm H}$ where $\mathfrak{g}_{\rm VT}$ and $\mathfrak{g}_{\rm H}$ are the Lie algebras of the isometries of the vector- and hypermultiplet moduli spaces respectively. The embedding tensor also splits into two components \cite{deWit_2011gk,LouisSmythTriendl_2012}.
Denoting by $k_a^{i}$ with $a =1 , \ldots, {\rm dim} \, \mathfrak{g}_{\rm VT}$ 
and $\tilde{k}_m{}^X$  with  $m =1 , \ldots, {\rm dim} \, \mathfrak{g}_{\rm H}$ the Killing vectors generating the vector- and hypermultiplet scalar manifold isometries respectively, the gauged generators can be written as
\begin{align}
 k^{i}_{\tilde{I}}(\phi)=\Theta_{\tilde{I}}{}^a k_a^{i}(\phi) \, ,
 \qquad \qquad
 \tilde{k}^X_{\tilde{I}}(q)=\Theta_{\tilde{I}}{}^m \tilde{k}_m^X(q).  
\end{align}

The two components of the embedding tensor $ \Theta_I{}^a $ and $ \Theta_I{}^m$ are thus constant
 $( n_{\textrm{VT}}+1)  \times  {\rm \dim} \, \mathfrak{g}_{\rm VT}$ and  $( n_{\textrm{VT}}+1)  \times  {\rm \dim} \, \mathfrak{g}_{\rm H}$  matrices, whose rank sets the dimension of the gauge group. 

\section{Exceptional Generalised Geometry}
\label{section_EGG}
Exceptional Generalised Geometry replaces the tangent bundle $TM$ with a larger bundle $E$ on $M_d$ whose fibres transform in a representation of the exceptional group $\Ex{d(d)}$. In doing so, the diffeomorphisms and the gauge symmetries of higher-dimensional SUGRA are unified as generalised diffeomorphisms on $E$. Then, one can similarly extend all conventional notions of differential geometry like tensors, connections, Lie derivatives and $G_S$-structures. For a more complete review, we refer to \cite{CoimbraStricklandWaldram_2014,AshmoreWaldram_2017}.

Compactifications of $11d$ supergravity on a $6d$ manifold $M$ are described by $\Ex{6(6)}\times \mathbb{R}^+$ exceptional geometry. The sections of the generalised tangent bundle $E$ transform in its $\mathbf{27}$ representation. Thanks to the embedding of the usual structure group $\GL{6}$ in this group, we decompose $E$:
\begin{align}
\label{generalised_tangent_bundle}
    E \simeq TM \oplus \wedge^2 T^{*}M \oplus \wedge^5 T^{*}M.
\end{align}
Its sections, called generalised vectors, can be described locally as sums of vectors, two-forms and five-forms on the internal manifold:
\begin{align}
    \label{generalised_vectors}
    V = v + \omega + \sigma.
\end{align}

Dual generalised vectors are sections of the dual bundle $E^*$. Generalised tensors are obtained by tensoring powers of $E$ and $E^*$. Most importantly, the generalised metric $G$ is a positive-definite, symmetric rank-2  section of the symmetric product $S^2 (E^*)$, encoding the internal components of all bosonic fields on $M$. Similarly to an ordinary metric, at any point on $M$ the generalised metric parameterises the coset
\begin{align}
    \label{generalised_metric_coset}
    G \in \frac{\Ex{d(d)}}{H_d} \overset{d=6}{=} \frac{\Ex{6(6)}}{\USp{8}/\mathbb{Z}_2},
\end{align}
where $H_d$ is the maximally compact subgroup of $\Ex{d(d)}$. For $d=6$, the generalised metric defines a reduced $\USp{8}/\mathbb{Z}_2$ structure. To account for the fermionic content of $11d$ supergravity, we also introduce spinors as sections of the spinor bundle $\mathcal{S}$, transforming in the spinorial representation of $\widetilde{H}_d$, the double cover of $H_d$. In our case, $\widetilde{H}_6 = \USp{8}$. 

We denote by $\ad \widetilde{F}$ the adjoint bundle, whose fibres transform in the adjoint of $\Ex{d(d)}$. The action of an infinitesimal generalised diffeomorphism is generated by the generalised Lie derivative along a generalised vector, which is defined as an adjoint $\Ex{d(d)}$ action \cite{CoimbraStricklandWaldram_2014b} by
\begin{align}
    \label{egg_dorfman}
    (L_V V')^M = V^N \partial_N V'^M - {(\partial \times_{\rm ad} V)^M}_N V'^N,
\end{align}
written in components in a standard atlas. Here, $\partial_M = \partial_m$ are viewed as sections of $E^*$ and we introduced the projection $\times_{\ad}: \, E^* \otimes E \to \ad \widetilde{F}$.

Extending the conventional definition, a generalised $G_S$-structure on $M$ is the reduction of the generalised structure group $\Ex{d(d)}$ to a subgroup $G_S$ and is defined by a set of non-vanishing $G_S$-invariant generalised tensors. The generalised metric already defines a $G_S = H_d$ structure: we will always look for further reductions. Given such a reduced structure, we define its torsion $T: \, \Gamma(E) \to \Gamma(\ad \widetilde{F})$ by its adjoint action on a generalised tensor $Q$:
\begin{align}
    \label{egg_torsion}
    (L^D_V - L_V) \, Q = T(V) \cdot Q,
\end{align}
where $D$ is any $G_S$-compatible connection\footnote{Given a $G_S$ structure with $G_S$-invariant generalised tensors $\{Q_i\}$, a generalised connection $D$ is $G_S$-structure compatible if $DQ_i = 0$ for all $Q_i$. Contrarily to conventional differential geometry, the conditions of being torsion-free and metric-compatible do not uniquely determine the connection. However, only certain projections of the action of the connection appear in supergravity, which can be shown to be unique \cite{CoimbraStricklandWaldram_2014b}.} and $L^D_V$ is the generalised Lie derivative computed with $D$. Then, the intrinsic torsion $T_{\rm int}$ is the portion of $T$ that is independent of the specific choice of $D$ and can be decomposed into representations of $G_S$.

\section{Consistent truncations}
Consider eleven-dimensional supergravity on a product space $X \times M$, where $X$ is a $D$-dimensional Lorentzian space and $M$ is a $d$-dimensional internal manifold in M-theory.

The supergravity fields \eqref{total_content} can be rearranged into generalised tensors as representations of $\GL{D,\mathbb{R}} \times \Ex{d(d)}$, reinterpreting the theory as a $D$-dimensional one with an infinite number of fields. The scalar degrees of freedom on $X$ are repackaged into a generalised metric, while the $\GL{D,\mathbb{R}}$ one-forms and vectors are sections of $E$. Finally, the two-forms are local sections of the weighted dual bundle $N\simeq \det T^*M \otimes E^*$\footnote{Any higher form-field would be dual to the scalar, vector or two-forms and thus would not introduce any new degree of freedom.}. Summarising,
\begin{align}
    \begin{split}
        \label{sugra_fields_repackage}
        \textrm{Scalars: }& \qquad G_{MN} (x,y)\in \Gamma(S^2 E^*),\\
        \textrm{Vectors: }& \qquad {\mathcal{A}_\mu}^M (x,y) \in \Gamma(T^* X \otimes E),\\
        \textrm{Two-forms: }& \qquad {\mathcal{B}_{\mu\nu}}^{MN} (x,y) \in \Gamma(\wedge^2 T^* X \otimes N),\\
    \end{split}
\end{align}
where $x$ and $y$ are coordinates on $X$ and $M$ respectively. The equations of motion and the supersymmetry variations are also rewritten according to the suitable representations of $\Ex{d(d)}$; the dynamics is completely set by the Levi-Civita connection on $X$ and a generalised connection $D$ on $M$.

The framework to study type IIB or M-theory consistent truncations to five dimensions with $\mathcal{N}=2$ supersymmetry was described in \cite{CassaniJossePetriniWaldram_2021}. Later, \cite{JosseMalekPetriniWaldram_2021} classified all such consistent truncations. It was found \cite{CassaniJossePetriniWaldram_2019} that a consistent truncation is obtained every time that a SUGRA theory is reduced on a manifold admitting a generalised $G_S$-structure with constant singlet intrinsic torsion, with $G_S \subset H_d$. The truncation is practically realised by expanding all supergravity fields on the invariant tensors under the structure group. The consistency of the truncation is then ensured by the fact that product of singlets will never source non-singlet representation. And also the intrinsic torsion being singlet no non-singlet representation can be sourced through the equations of motion.

From the data of the $G_S$-structure, we can determine the number of fields of the truncated theory and all the possible gaugings. For instance, the scalars of the truncated theory are given by the $G_S$-singlets in $G_{MN}$; they are singlet deformations of the structure, modulo the deformations that do not affect the metric:
\begin{align}
    \label{trunc_scalars_coset}
    h^{\tilde{I}}(x) \in \mathcal{M} = \frac{{\rm C}_{\Ex{d(d)}} (G_S)}{{\rm C}_{H_d} (G_S)} =: \frac{\mathcal{G}}{\mathcal{H}}.
\end{align}
The vectors are determined by the $G_S$-invariant generalised vectors $\{K_{\tilde{I}}\}$, which span a vector space $\mathcal{V}$:
\begin{align}
    \label{trunc_vectors}
    {\mathcal{A}_\mu}^M (x,y) = {A_\mu}^{\tilde{I}} (x) {K_{\tilde{I}}}^M \in \Gamma(T^*X \otimes \mathcal{V}).
\end{align}
The vectors $K_{\tilde{I}}$ generate all symmetries of the reduced theory. The two-form is expanded on the singlets of the bundle $N$. The $G_S$-structure also determines the embedding tensor \cite{Samtleben_2008, Trigiante_2016} and thus the gaugings of the reduced theory in terms of $T_{\rm int}$. As the $G_S$-structure has only singlet intrinsic torsion, the generalised Lie derivative of the invariant generalised tensors $\{Q_i\}$ along any of the $K_{\tilde{I}}$ is
\begin{align}
    \label{trunc_dorfman_on_tensors}
    L_{K_{\tilde{I}}} Q_i = - T_{\rm int}(K_{\tilde{I}}) \cdot Q_i,
\end{align}
where $T_{\rm int}$ defines a linear map from the space of $G_S$-singlet vectors $\mathcal{V}$ to the Lie algebra of $\mathcal{G}$\footnote{$T_{\rm int}$ maps $\mathcal{V}$ to the $G_S$-singlets in the adjoint bundle. Thus, $T_{int} (K_{\tilde{I}})$ corresponds to the elements in the adjoint that commute with $G_S$, namely the Lie algebra of $\mathcal{G}$, given in \eqref{trunc_scalars_coset}. $\mathcal{G}$ is the subgroup of the isometry group of the scalar manifold that can be gauged in the truncated theory.}; we can identify it with the embedding tensor of the truncated theory.

\subsection{Five-dimensional $\mathcal{N}=2$ truncations}
In five-dimensional $\mathcal{N}=2$ supergravity, the R-symmetry group is $\SU{2}$. Because of how this group embeds in $\Ex{6(6)}$, the largest structure group preserving $\mathcal{N}=2$ supersymmetry occurs for $G_S=\USp{6}$:
\begin{align}
\begin{split}
    \label{n2_usp8_decomposition}
    \USp{8} &\supset \SU{2}_R \times \USp{6},\\
    \mathbf{8} &= (\mathbf{6},\mathbf{1}) \oplus (\mathbf{1},\mathbf{2}),
\end{split}    
\end{align}
where the two $\USp{6}$ singlets give the $\SU{2}_R$ doublet of supersymmetry parameters.

Decomposing the fundamental $\mathbf{27}$ and the adjoint $\mathbf{78}$ representations of $\Ex{6(6)}$ under \eqref{n2_usp8_decomposition}, we find one singlet generalised vector $K \in \Gamma(E)$ of positive norm with respect to the $E_{6(6)}$ cubic invariant\footnote{This is six times the cubic invariant given in \cite{AshmoreWaldram_2017}}, 
\begin{align}
    \label{n2_cubic}
    c(K,K,K)=6 \, \kappa^2 > 0,
\end{align}
with $\kappa \in \Gamma((\det T^*M)^{1/2})$, and a $\SU{2}$ triplet of adjoint tensors $J_\alpha \in \Gamma(\ad \widetilde{F})$, such that 
\begin{align}
    \label{n2_H_structure}
    [J_\alpha, J_\beta] = 2  \epsilon_{\alpha \beta \gamma} J_\gamma, \qquad \tr(J_\alpha J_\beta)= -  \delta_{\alpha \beta}.
\end{align}

The global, positive-normed vector $K$ is called the vector-multiplet structure, or V-structure. The triplet $J_\alpha$ is called the hypermultiplet structure, or H-structure. Together, when they satisfy the compatibility condition
\begin{align}
    \label{HV_compatibility}
    J_\alpha \cdot K = 0
\end{align}
$K$ and $J_\alpha$ define an HV-structure. For $D=5$, this corresponds to having a $\USp{6}$-structure\footnote{The generalised metric is given in terms of the $G_S$-invariant tensors as \cite{CassaniJossePetriniWaldram_2021}
\begin{align}
    \label{generalised_metric}
    G(V,V) = 3 \bigg( 3 \frac{c(K,K,V)^2}{c(K,K,K)^2} - 2 \frac{c(K,V,V)}{c(K,K,K)} + 4 \frac{c(K,J_3 \cdot V, J_3 \cdot V)}{c(K,K,K)} \bigg).
\end{align}}.

The reduced theory with such an HV-structure is minimal $\mathcal{N}=2$ SUGRA, and is rigid up to a scaling of $\kappa^2$, meaning that it has no room for any vector- or hypermultiplets. To allow for any extra multiplets, it is necessary to reduce the structure group further than $\USp{6}$. A reduced $G_S \subset \USp{6}$ structure leads naturally to a moduli space of $G_S$-invariant HV-structures. A smaller structure group admits more $G_S$-singlets in the decomposition of the $\textbf{27}$ and the $\textbf{78}$, though no new singlets may arise in the decomposition of the spinorial representation of $\USp{8}$ since we want only $\mathcal{N}=2$ supersymetry. A generic $G_S \subset \USp{6}$ corresponding to $\mathcal{N}=2$ SUSY in $D=5$ is defined by the set of $G_S$-invariant generalised vectors $K_{\tilde{I}}$ and generalised adjoint tensors $J_A$, $A=1,\dots,\dim\mathcal{H}$, also satisfying the compatibility condition
\begin{align}
    \label{KIJA_compatibility_constraint}
    J_A \cdot K_{\tilde{I}} = 0, \qquad \forall \, {\tilde{I}}, \, \forall A.
\end{align}
This is precisely why the scalar manifold $\mathcal{M}$ splits according to \eqref{total_scalar_mfld}. The $J_A$ generate the isometry group $\mathcal{H}$ of the hypermultiplet scalar manifold $\mathcal{M}_{\rm H}$, so that
\begin{align}
    \label{JA_algebra}
    [J_A,J_B]={f_{AB}}^C J_C,
\end{align}
${f_{AB}}^C$ denoting the structure constants of $\mathcal{H}$. We also normalise the $K_{\tilde{I}}$ and $J_A$ by
\begin{align}
    \label{normalisation_KI_JA}
    c(K_{\tilde{I}},K_{\tilde{J}},K_{\tilde{K}})= 6 \kappa^2 C_{\tilde{I}\tilde{J}\tilde{K}},\qquad \tr(J_A J_B) = \eta_{AB},
\end{align}
with $C_{\tilde{I}\tilde{J}\tilde{K}}$ a constant, symmetric tensor and $\eta_{AB}$ a diagonal matrix, with -1 and +1 entries corresponding to compact and non-compact generators of $\mathcal{H}$.

The $G_S$-structure determines the field content of the truncated theory: the singlet generalised vectors are in one-to-one correspondence with the truncated theory vectors. Out of the tensors in this structure, it is always possible to define an HV-structure, building an appropriate set $\{K,J_\alpha \}$. Any such HV-structure can be related to another by an $\Ex{6(6)} \times  \mathbb{R}^{+}$ transformation. The $\mathbb{R}^{+}$ factor just rescales the density $\kappa^2$.

The V-structure moduli space $\mathcal{M}_{\rm V}$ corresponds to all deformations of $K$ that leave $J_\alpha$ invariant and is obtained by writing $K$ as a linear combination of the invariant vectors, $K_{\tilde{I}}$
\begin{align}
    \label{K_hIKI}
    K = h^{\tilde{I}} K_{\tilde{I}},
\end{align}
where the scalars $h^{\tilde{I}}$ must satisfy \eqref{cubic_hypersurface} because of the condition \eqref{n2_cubic}. $\mathcal{M}_{\rm V}$ is the hypersurface defined by these scalars. The metric on $\mathcal{M}_{\rm V}$ is given by $a_{\tilde{I}\tilde{J}} = \tfrac{1}{3} G(K_{\tilde{I}},K_{\tilde{J}})$ and reproduces the five-dimensional expression \eqref{ambient_space_metric}.

The H-structure moduli space $\mathcal{M}_{\rm H}$ corresponds to all deformations of $J_\alpha$ preserving $K$, i.e. all the highest root $\mathfrak{su}_2$ subalgebras in the algebra spanned by the $J_A$. This space is  quaternionic-K\"ahler. We can always build a triplet of ``dressed'' generalised tensors $J_\alpha$ as the combination
\begin{align}
    \label{J_mAJA}
    J_\alpha = m^A_\alpha J_A.
\end{align}
We can immediately plug this linear combination in the normalisation of the H-structures \eqref{KIJA_compatibility_constraint} to get a useful identity for the embedding matrices $m^A_\alpha$:
\begin{align}
    \label{identity_m}
    m^A_\alpha m^B_\beta \eta_{AB} = \delta_{\alpha \beta}.
\end{align}
The two sets of deformations are independent. If we allow them to depend on the external spacetime coordinates, these deformations coincide exactly with the scalar fields in the low-dimensional theory. 

\subsection{Intrinsic torsion}
\label{section:intrinsic_torsion}
For $\mathcal{N}=2$ SUSY, the gauging can be expressed in term of the embedding tensor, split into $({\Theta_{\tilde{I}}}^a,{\Theta_{\tilde{I}}}^A)$ to reflect the split of the isometry algebra $\mathfrak{g}_{\rm iso} = \mathfrak{g}_{\rm V} \oplus \mathfrak{g}_{\rm H}$ as mentioned earlier.

The intrinsic torsion is encoded in the differential expressions
\begin{align}
\begin{split}
    \label{intrinsic_torsion_HV}
    L_{K_{\tilde{I}}} K_{\tilde{J}} &= \Theta_{\tilde{I}} \cdot K_{\tilde{J}} = {\Theta_{\tilde{I}}}^a {(t_a)_{\tilde{J}}}^{\tilde{K}} K_{\tilde{K}} =: {t_{\tilde{I}\tilde{J}}}^{\tilde{K}} K_{\tilde{K}},\\
    L_{K_{\tilde{I}}} J_A &= \Theta_{\tilde{I}} \cdot J_A = [{\Theta_{\tilde{I}}}^B J_B, J_A] = {\Theta_{\tilde{I}}}^B {f_{BA}}^C J_C =: {p_{{\tilde{I}}A}}^B J_B,
\end{split}
\end{align}
where $t_a$ are the generators of $\Lie \, \mathcal{G}$ acting on $\mathcal{V}$, ${t_{\tilde{I}\tilde{J}}}^{\tilde{K}}$ are the structure constants of $\mathfrak{g}_{\rm gauge}$ and ${f_{BA}}^C$ the ones of the algebra acting on the hypers. We refer to \cite{JosseMalekPetriniWaldram_2021} for more details. 

In order to have a consistent truncation we ask for the intrinsic torsion to be singlet and constant. The matrices ${t_{\tilde{I}\tilde{J}}}^{\tilde{K}}$ and ${p_{\tilde{I}A}}^B$ must be constant and for the intrinsic torsion to be singlet we additionally ask that
\begin{align}
    \label{moment_map_constraint_singlet_intrinsic}
    \int_M \kappa^2 \tr(J_A L_W J_B) = 0, 
\end{align}
for all vectors $W$ orthogonal to the basis $\{K_{\tilde{I}}\}$, by which we mean  
\begin{align}
c(K_{\tilde{I}},K_{\tilde{J}},W) = 0\,.
\end{align}
Knowing the intrinsic torsion of every HV-structure in the family fixes the intrinsic torsion of the total $G_S$-structure. 

\subsection{Exceptional Sasaki-Einstein structures}
In \cite{AshmorePetriniWaldram_2016}, the exceptional generalised geometry analogue of a Sasaki–Einstein
structure was defined, corresponding to an $\mathcal{N}=2$ AdS background with generic fluxes. An exceptional Sasaki-Einstein (ESE) structure is a HV-structure $\{J_\alpha, K\}$ ($\alpha=1,2,3$) satisfying 

\begin{equation}
\begin{aligned}
\label{ESEHVd5}
&\mu_\alpha (V) = \lambda_\alpha \gamma(V) \quad \forall V \in \Gamma(E), \\
&L_K K=0,\\
&L_K J_\alpha = \epsilon_{\alpha\beta\gamma} \lambda_\beta J_\gamma,
\end{aligned}
\end{equation}
where $\lambda_\alpha$ are real constants, whose norm norm is related to the AdS inverse radius, $m$. We also define 
\begin{align}
    \label{moment_map_triplet}
    \mu_\alpha (V) = -\frac{1}{2} \epsilon_{\alpha \beta \gamma} \int_M \kappa^2 \tr(J_\beta L_V J_\gamma),
\end{align}
the moment maps for the action of the generalised diffeomorphisms on the space of H-structures, and
\begin{align}
    \label{gamma_function}
    \gamma(V) = \int_M \kappa^2 c(V,K,K).
\end{align}
As in \cite{AshmorePetriniWaldram_2016}, we can always use the global $\SU{2}_R$ symmetry to set $\lambda_1 = \lambda_2 =0$. Then, the only unbroken part of the R-symmetry is a $\Uni{1}_R$ factor preserving $\lambda_3=3m$.

\section{Deriving the BPS equations}

In this section, we will re-express the ESE constraints \eqref{ESEHVd5} using the truncation data and the gaugings mentioned above. Recall the expansions of the HV-structure $\{K,J_\alpha\}$ on the bases $\{K_{\tilde{I}}, J_A\}$:
\begin{align}
    K = h^{\tilde{I}} K_{\tilde{I}} \quad \text{and} \quad J_\alpha = m^A_\alpha J_A.
\end{align}
Starting with the second equation of \eqref{ESEHVd5}, we find:
\begin{align}
    \label{LKK_ET_1}
    & 0 = L_K K = L_{h^{\tilde{I}} K_{\tilde{I}}} (h^{\tilde{J}} K_{\tilde{J}}) = h^{\tilde{I}} h^{\tilde{J}} L_{K_{\tilde{I}}} K_{\tilde{J}} = h^{\tilde{I}} h^{\tilde{I}} {t_{\tilde{I}\tilde{J}}}^{\tilde{K}} K_{\tilde{K}}, 
\end{align}
where the last step comes from \eqref{intrinsic_torsion_HV}. Remark that the embedding coordinates $h^{\tilde{I}}$ go through the Dorfman derivative, since they do not depend on the internal coordinates. To conclude, we consider the action of the cubic invariant $C$ on this expression and two more elements from the V-structure:
\begin{align}
    \label{LKK_ET_2}
    C(K_{\tilde{L}},K_{\tilde{M}}, h^{\tilde{I}} h^{\tilde{I}} {t_{\tilde{I}\tilde{J}}}^{\tilde{K}} K_{\tilde{K}}) = h^{\tilde{I}} h^{\tilde{I}} {t_{\tilde{I}\tilde{J}}}^{\tilde{K}} C_{{\tilde{L}}{\tilde{M}}{\tilde{K}}} = 0.
\end{align}

Moving onto the third ESE constraint from \eqref{ESEHVd5}, we act with the $\mathfrak{e}_{6(6)}$ Killing form on both sides of the equation, tracing with another adjoint element $J_D$:
\begin{align}
\begin{split}
    \label{LKJ_ET_2}
    & \tr(h^{\tilde{I}} m^A_\alpha {p_{{\tilde{I}}A}}^B J_B,\, J_D) = \tr(\epsilon_{\alpha \beta \gamma} \lambda_\beta m^C_\gamma J_C, \, J_D),\\
   \Rightarrow\, & h^{\tilde{I}} m^A_\alpha {p_{{\tilde{I}}A}}^B \eta_{BD} = \epsilon_{\alpha \beta \gamma} \lambda_\beta m^C_\gamma \eta_{CD},\\
    \Rightarrow\, & h^{\tilde{I}} m^A_\alpha {p_{{\tilde{I}}A}}^B = \epsilon_{\alpha \beta \gamma} \lambda_\beta m^B_\gamma.
\end{split}
\end{align}
Lastly, we consider the moment map equation, the first of the ESE constraints from \eqref{ESEHVd5}, using \eqref{moment_map_triplet}:
\begin{align}
\begin{split}
    \label{MM_ET_1}
    &-\frac{1}{2} \epsilon_{\alpha \beta \gamma} \int_M \kappa^2 \tr(J_\beta, L_V J_\gamma) = \lambda_\alpha \int_M \kappa^2 c(V,K,K),\\
   \Rightarrow\,  & -\frac{1}{2} \epsilon_{\alpha \beta \gamma} m^A_\beta m^B_\gamma \int_M \kappa^2 \tr(J_A, L_V J_B) = \lambda_\alpha h^{\tilde{I}}  h^{\tilde{J}} \int_M \kappa^2 c(V, K_{\tilde{I}},  K_{\tilde{J}}).
\end{split}
\end{align}
Next, we expand the arbitrary generalised vector $V \in \Gamma(E)$ on the basis of generalised vectors $K_{\tilde{I}}$ and their orthogonal ones. Only the singlet part will contribute due to \eqref{moment_map_constraint_singlet_intrinsic}. Hence, $v^{\tilde{K}} K_{\tilde{K}} \in V$ gives the only non-zero contribution to the equation above. Since $L_V J_B = -T_{\rm{int}} (V) \cdot J_B = v^{\tilde{K}} {p_{\tilde{K}B}}^C J_C$, we therefore have:
\begin{align}
\begin{split}
    \label{MM_ET_2}
    & -\frac{1}{2} \epsilon_{\alpha \beta \gamma} m^A_\beta m^B_\gamma  {p_{\tilde{K}B}}^C \int_M \kappa^2 v^{\tilde{K}} \tr(J_A, J_C)= \lambda_\alpha h^{\tilde{I}}  h^{\tilde{J}}  \int_M \kappa^2 c(v^{\tilde{K}} K_{\tilde{K}}, K_{\tilde{I}},  K_{\tilde{J}}),\\
   &\Rightarrow\,   -\frac{1}{2} \epsilon_{\alpha \beta \gamma} m^A_\beta m^B_\gamma p_{\tilde{K}BA} = \lambda_\alpha h_{\tilde{K}}.
\end{split}
\end{align}
We can now summarise the full translation of the ESE constraints \eqref{ESEHVd5}: 
\begin{align}
\begin{split}
\label{ESE_to_ET_formalism}
& -\frac{1}{2} \epsilon_{\alpha \beta \gamma} m^A_\beta m^B_\gamma p_{\tilde{K}BA} = \lambda_\alpha h_{\tilde{K}}, \\
& h^{\tilde{I}} h^{\tilde{J}} {t_{\tilde{I}\tilde{J}}}^{\tilde{K}} C_{{\tilde{L}}{\tilde{M}}{\tilde{K}}} = 0, \\
& h^{\tilde{I}} m^A_\alpha {p_{{\tilde{I}}A}}^B = \epsilon_{\alpha \beta \gamma} \lambda_\beta m^B_\gamma. 
\end{split}
\end{align}
Next, we claim that these equations correspond exactly to the BPS equations for 5d $\mathcal{N}=2$ supergravity presented in \cite{LouisMuranaka_2016}. Indeed, from \cite{JosseMalekPetriniWaldram_2021}, the prepotential can be written as
\begin{align}
    \label{prepotential_ET}
    P^{\alpha}_{\tilde{I}} = \frac{1}{g} \epsilon^{\alpha \beta \gamma} m^B_{\beta} m^C_{\gamma} p_{\tilde{I}CB}.
\end{align}
Thus, the first equation of \eqref{ESE_to_ET_formalism} takes the form:
\begin{align}
    P_{\alpha,\,\tilde{I}} = -\frac{2}{g}  \lambda_\alpha h_{\tilde{I}}.
\end{align}
Contracting with $h^{\tilde{I}}$ and $h^{\tilde{I}}_i$, we can obtain the following equations:
\begin{align}
    h^{\tilde{I}} P_{\alpha,\,\tilde{I}} &= -\frac{2}{g}  \lambda_\alpha, \\
    h^{\tilde{I}}_i P_{\alpha,\,\tilde{I}} &=0.  
\end{align}
For the next equations we will need the following identities coming from \cite{JosseMalekPetriniWaldram_2021},
\begin{align} 
h^{\tilde{J}} {t_{\tilde{I}\tilde{J}}}^{\tilde{K}}&=k^i_{\tilde{I}}  h^{\tilde{K}}_i, \\
m^A_\alpha {p_{{\tilde{I}}A}}^B &=\tilde{k}^X_{\tilde{I}} \partial_X m^B_\alpha.
\end{align}
Moving onto the second equation:
\begin{align}
\begin{split}
    h^{\tilde{I}} k^i_{\tilde{I}}  h^{\tilde{K}}_i  C_{{\tilde{L}}{\tilde{M}}{\tilde{K}}} = 0, \\
    h^{\tilde{I}} k^i_{\tilde{I}}  h^{\tilde{K}}_i  C_{{\tilde{L}}{\tilde{M}}{\tilde{K}}} h^{\tilde{L}}h^{\tilde{M}}_j = 0, \\
    h^{\tilde{I}} k^i_{\tilde{I}} g_{ij} = 0.
\end{split}
\end{align}
Finally, the last equation can be transformed as:
\begin{align}
\begin{split}
h^{\tilde{I}} \tilde{k}^X_{\tilde{I}} \partial_X m^A_\alpha &= \epsilon_{\alpha \beta \gamma} \lambda_\beta m^A_\gamma, \\
h^{\tilde{I}} \tilde{k}^X_{\tilde{I}} \partial_X m^A_\alpha \partial_Y m^B_\sigma\eta_{AB}\delta^{\alpha\sigma} &= \epsilon_{\alpha \beta \gamma} \lambda_\beta m^A_\gamma \partial_Y m^B_\sigma \eta_{AB}\delta^{\alpha\sigma},\\
h^{\tilde{I}} \tilde{k}^X_{\tilde{I}} f_{XY}&=0,
\end{split}
\end{align}
where the RHS is zero because of \eqref{identity_m} and $f_{XY}=\partial_X m^A_\alpha \partial_Y m^B_\sigma\eta_{AB}\delta^{\alpha\sigma}$ is an invertible matrix. Hence, we obtain the following BPS equation:
\begin{align}
h^{\tilde{I}} \tilde{k}^X_{\tilde{I}}=0.
\end{align}
Our transformed set of equations is given by
\begin{align}
h^{I} P_{\alpha,\,I} &= -\frac{2}{g}  \lambda_\alpha \, , \quad h^{I}_i P_{\alpha,\,I} &=0 \, , \quad  h^{\tilde{I}} \tilde{k}^X_{\tilde{I}}=0 \, , \quad h^{\tilde{I}} k_{i,\,\tilde{I}}= 0.
\end{align}
As promised, this is exactly the set presented in \cite{LouisMuranaka_2016}. For our vacua study it is actually more convenient to use the set of equations \eqref{ESE_to_ET_formalism}.

\section{Looking for AdS vacua}
In every consistent truncation of the classification determined in \cite{JosseMalekPetriniWaldram_2021} and for every gauging thereof, we have all the ingredients to evaluate the BPS equations above to check whether there is a AdS vacuum with $\mathcal{N}=2$. Not all gaugings need to be checked: as found by \cite{LouisMuranaka_2016}, the unbroken gauge group of the $\textrm{AdS}_5$ vacua must contain an $\Uni{1}_R$ factor. Thus, we can immediately discard any listed gauging that does not feature this factor. We recall all surviving cases in Table \ref{t:summary}. We will now only describe the consistent truncations that were found to admit an AdS vacuum. More details are available in \cite{Merenda_2023}. 

\subsection{$n_{\rm VT}=0$ and $n_{\rm H}=0$ with $\Uni{1}_R$ gauging}
This very simple case of truncation obtained with an $\USp{6}$ structure contains an AdS vacuum. It actually corresponds to minimal 5d supergravity with only the supergravity multiplet and thus only one scalar, $H_0$. The embedding tensor is chosen to be\footnote{This is a direct consequence of choosing $\lambda_1 = \lambda_2 =0$ and having $\U(1)_R$ as a gauge group.}
\begin{align}
p_{0\, AB}=\begin{pmatrix}
\varepsilon_{ab} & 0 \\
0 & 0 
\end{pmatrix}  \quad \text{and} \quad m_\alpha^A=\delta_\alpha^A, 
\end{align}
where $A\equiv \alpha=1,2,3$ and $a=1,2$. The BPS equations reduce to 
\begin{align}
H_0=-\frac{1}{2}\lambda_3. 
\end{align}

\subsection{$n_{\rm VT}=0$ and $n_{\rm H}=2$ with $\Uni{1}\times\Uni{1}_R$ gauging}
In this case, there are eight scalars coming from the hyper-sector: $\{a,y,h,U,Z_1,Z_2,Z_3,Z_4\}$. The way we worked out the relevant $m_\alpha{}^A$ matrices and $p_{0\, AB}$ tensor has been detailed in Appendix \ref{app__g22_construction}. Plugging those data in the BPS equations, the only solution we find is:
\begin{align}
H_0=-\frac{1}{2} \lambda_3 \,, \quad a=y=h=U=Z_1=Z_2=Z_3=Z_4=0.
\end{align}
This solution is the same as the one in the minimal $\USp{6}$ structure truncation, in the sense that the extra scalars brought by the two extra hypermultiplets do not achieve a new minimum on the scalar manifold.

\subsection{$n_{\rm VT}=0$ and $n_{\rm H}=1$ with $
\Uni{1}_R$ gauging}
In this case, there are four scalars from the hyper-sector, coming from the construction presented in Appendix F of \cite{CassaniJossePetriniWaldram_2021}.
Plugging those data in the BPS equations, we find the solution:
\begin{align}
H_0=-\frac{1}{2} \lambda_3 \,, \quad \theta_1=\theta_2=\varphi=\xi=0.
\end{align}
Like in the previous case, no new vacuum is obtained by the extra scalars and we recover again the previous minimal solution.

\subsection{$n_{\rm VT}=1$ and $n_{\rm H}=1$ with $\Uni{1}\times\Uni{1}_R$ gauging}

The consistent truncation to this 5d gauged supergravity has been constructed in \cite{Szepietowski_2012tb,CassaniJossePetriniWaldram_2021}. The uplift of this $\AdS_5$ $\mathcal{N}=2$ corresponds to \cite{Bah_2011vv}. Most of the objects ($C_{{\tilde{L}}{\tilde{M}}{\tilde{K}}}\,,\,m^A_\alpha\,,\,\ldots$) were already worked out. Using this truncation data, encoding properly the gauging and plugging it in the BPS equations, we find that the only possible solution is the family of solutions that precisely corresponds to the BBBW solutions.

\subsection{$n_{\rm VT}=4$ and $n_{\rm H}=1$ with $\SO{3}\times\Uni{1}_R$ gauging}
The consistent truncation to this 5d gauged supergravity has been constructed in \cite{Faedo_2020,CassaniJossePetriniWaldram_2021}. The uplift of this $\AdS_5$ $\mathcal{N}=2$ corresponds to \cite{Maldacena_2000mw}. As for the previous case part of the truncation data can be reused and once we implemented the gauging correctly we plug it in the BPS equations. In the end the only  solution is the Maldacena-N\'u\~{n}ez solution.

\section{Summary and conclusions}
In this letter we briefly reviewed $\mathcal{N}=2$ $D=5$ SUGRA, the exceptional geometry associated with it and how to build consistent truncations within this framework. We made explicit how the conditions for exceptional Sasaki-Einstein backgrounds are linked to the five-dimensional $\mathcal{N}=2$ supersymmetric equations. Then, the vacua investigation led us to find the minimal solution and recover the known solutions within the truncations containing the Maldacena-N\'u\~{n}ez solution \cite{Maldacena_2000mw} and the family of BBBW solutions \cite{Bah_2011vv}. This study seems to suggest that, except from those known solutions, there is no other $\AdS_5$ solution with $\mathcal{N}=2$ supersymmetry coming from $\mathcal{N}=2$ consistent truncations of eleven or ten dimensional supergravity down to five dimensions. Put differently, it means that for all the $\AdS_5$ $\mathcal{N}=2$ backgrounds that could potentially be uplifted, an uplift has already been found. Notice that by construction solutions with supersymmetry enhancement are not included under our scope. An interesting future direction would be to understand what are the implication of such conclusion on the dual field theory side. Another direction under investigation is to produce the same kind of classification and AdS scan but in other dimensions. 

\begin{table*}[h]
    
\makebox[\textwidth][c]{
\begin{tabular}{|c|c|c|c|}
\hline
$(n_{\rm VT}, n_{\rm H})$ & $G_S, \, G_{\rm iso},\, \mathcal{M}$ & $G_{\rm gauge}$ & $n_{\rm T}$  \\
\hline
(1,0)& 
\miniworld{$G_S=\SU{2}\times\Spin{5}$\\
$G_{\rm iso} = \SU{2}_R$\\
$\mathcal{M}=\mathbb{R}^+$}
& $\Uni{1}_R$
&$-$\\
\hline
(2,0)& 
\miniworld{$G_S=\SU{2}\times\Spin{4}$\\
$G_{\rm iso} = \SU{2}_R \times \SO{1,1} \times \mathbb{R}^+$\\
$ \mathcal{M}=\mathbb{R}^+ \times \SO{1,1}$}
& $\Uni{1}_R$
&$-$\\
\hline
(3,0)& 
\miniworld{$G_S=\SU{2}\times\Spin{3}$\\
$G_{\rm iso} = \SU{2}_R \times \SO{2,1} \times \mathbb{R}^+$\\
$ \mathcal{M}=\mathbb{R}^+ \times \frac{\SO{2,1}}{\SO{2}}$}
& $\SO{2,1} \times \Uni{1}_R$
&$-$\\
\hline
(4,0)& 
\miniworld{$G_S=\SU{2}\times\Spin{2}$\\
$G_{\rm iso} = \SU{2}_R \times \SO{3,1} \times \mathbb{R}^+$\\
$ \mathcal{M}=\mathbb{R}^+ \times \frac{\SO{3,1}}{\SO{3}}$}
& \centering
\miniworld{
$\SO{2,1} \times \Uni{1}_R$, $\SO{3} \times \Uni{1}_R$,
$\ISO{2} \times \Uni{1}_R$ \\
\vspace{0.1cm}
$\SO{2} \times \Uni{1}_R$, \, $\SO{1,1} \times \Uni{1}_R$}   
& \tinyworld{$-$ \\ \vspace{0.4cm} $\,\,\,2$ } 
\\
\hline
(6,0)& 
\miniworld{$G_S=\SU{2}\times\mathbb{Z}_2$\\
$G_{\rm iso} = \SU{2}_R \times \SO{5,1} \times \mathbb{R}^+$\\
$ \mathcal{M}=\mathbb{R}^+ \times \frac{\SO{5,1}}{\SO{5}}$}
& \centering
\miniworld{
$\SO{3} \times \SO{2,1} \times \Uni{1}_R$\\
\vspace{0.1cm}
$\SO{2,1} \times \Uni{1} \times \Uni{1}_R$, $\SO{3} \times \SO{2} \times \Uni{1}_R$, $\SO{3} \times \SO{1,1}\times \Uni{1}_R$, $\ISO{2}\times \Uni{1}\times \Uni{1}_R$\\
\vspace{0.1cm}
$\Uni{1}\times \Uni{1}_R$, $\SO{1,1} \times \Uni{1}_R$}     
&\tinyworld{\vspace{0.2cm}$-$ \\\vspace{0.7cm} $2$ \\\vspace{0.6cm} $\,\,4$}
\\     
\hline
(5,0)& 
\miniworld{$G_S=\SU{2}$\\
$G_{\rm iso} = \SU{2}_R \times \SL{3,\mathbb{R}}$\\
$ \mathcal{M}=\tfrac{\SL{3,\mathbb{R}}}{\SO{3}}$}
& \centering
$\SL{2,\mathbb{R}} \times \Uni{1}_R$     
& $2$ \\
\hline
(8,0)& 
\miniworld{$G_S=\Uni{1}$\\
$G_{\rm iso} = \SU{2}_R \times \SL{3,\mathbb{C}}$\\
$ \mathcal{M}=\tfrac{\SL{3,\mathbb{C}}}{\SU{3}}$ }
& \centering
\miniworld{$\SU{3} \times \Uni{1}_R, \, \SU{2,1} \times \Uni{1}_R$\\
\vspace{0.1cm}
$\SU{2} \times \Uni{1}_R, \, \SU{1,1}\times\Uni{1}_R$\\
\vspace{0.1cm}
$\Uni{1}\times\Uni{1}_R$}     
&\tinyworld{$-$ \\\vspace{0.15cm} $4$ \\\vspace{0.15cm} $\,\,6$}  \\
\hline
(14,0)&  
\miniworld{$G_S=\mathbb{Z}_2$\\
$G_{\rm iso} = \SU{2}_R \times \SUstar{6}$\\
$ \mathcal{M}=\tfrac{\SUstar{6}}{\USp{6}}$ }
& \centering
\miniworld{$\SU{3} \times \Uni{1}_R, \, \SU{2,1} \times \Uni{1}_R$  \\
\vspace{0.1cm}
$\SU{2} \times \Uni{1}_R$\\
\vspace{0.1cm}
$\Uni{1}\times\Uni{1}_R$}     
&\tinyworld{$6$ \\\vspace{0.15cm} $10$ \\\vspace{0.15cm} $\,\,12$}  \\
\hline
(0,1)&
\miniworld{$G_S=\SU{3}$\\
$G_{\rm iso} = \SU{2,1}$\\
$ \mathcal{M}=\tfrac{\SU{2,1}}{{\rm S}(\Uni{2} \times \Uni{1})}$}
& \centering
$\Uni{1}_R$      
& $-$ \\  
\hline
(0,2)& 
\miniworld{$G_S=\SO{3}$\\
$G_{\rm iso} = \Gx{2(2)}$\\
$ \mathcal{M}=\tfrac{\Gx{2(2)}}{\SO{4}}$}
& \centering
$\Uni{1}_R$  
& $-$  \\  
\hline
(1,1)& 
\miniworld{$G_S=\SU{2} \times \Uni{1}$\\
$G_{\rm iso} = \SU{2,1} \times \SO{1,1} \times \mathbb{R}^+$\\
$ \mathcal{M}=\mathbb{R}^+ \times \tfrac{\SU{2,1}}{{\rm S}(\Uni{2} \times \Uni{1})}$}
& \centering
$\mathbb{R}^+ \times \Uni{1}_R $      
& $-$ \\   
\hline
(4,1)& 
\miniworld{$G_S=\Uni{1}$\\
$G_{\rm iso} = \SU{2,1} \times \SO{3,1} \times \mathbb{R}^+$\\
$ \mathcal{M}=\mathbb{R}^+ \times \tfrac{\SO{3,1}}{\SO{3}} \times\tfrac{\SU{2,1}}{{\rm S}(\Uni{2} \times \Uni{1})}$}
& \centering
\miniworld{$\SO{2,1} \times \mathbb{R}^+ \times \Uni{1}_R $, $\SO{3} \times \mathbb{R}^+ \times \Uni{1}_R $, $\ISO{2} \times \mathbb{R}^+ \times \Uni{1}_R $\\
\vspace{0.1cm}
$\SO{2}\times \mathbb{R}^+ \times \Uni{1}_R$, $\SO{1,1}\times \mathbb{R}^+ \times \Uni{1}_R $}     
&\tinyworld{$-$\\ \vspace{0.9cm}$\,\,\,2$  }  \\
\hline
\end{tabular}}
\caption{List of all possible consistent truncations containing an AdS vacua, with $n_{\rm VT}$ vector/tensor multiplets and $n_{\rm H}$ hypermultiplets. In the second column, we display the required $G_S \subset \Ex{6(6)}$ structure group, the isometry group $G_{\rm iso}$ and the associated scalar manifold $\mathcal{M}$. In the third column, we show the allowed gaugings $G_{\rm gauge}$ of $G_{\rm iso}$ after the filter of \cite{LouisMuranaka_2016}, reporting the number of vectors that are dualised to tensors in each case.} 

\label{t:summary}
\end{table*}

\section*{Acknowledgements}
We would like to thank Michela Petrini, Davide Cassani, Emanuel Malek and Nikolay Bobev for useful discussions. GJ is supported by the Deutsche Forschungsgemeinschaft (DFG, German Research Foundation) via the Emmy Noether program ``Exploring the landscape of string theory flux vacua using exceptional field theory'' (project number 426510644).

\appendix

\section{$\tfrac{\Gx{2(2)}}{\SU{2}\times \SU{2}}$ construction}
\label{app__g22_construction}
The $\SU{2}$ triplet embedding matrices $m_\alpha^A$ and the ${p_{0\, A}}^B$ tensor for the case with $n_{\rm VT}=0$, $n_{\rm H}=2$ have been worked out using the conventions and the details of the construction of the $\tfrac{\Gx{2(2)}}{\SU{2}\times \SU{2}}$ coset from \cite{Fre_2018}. 

The generators $j_\alpha$ of the $\SU{2}$ subalgebras inside $\Gx{2(2)}$ are given in \cite{Fre_2018}-(7.7.23) as linear combinations of the $\Gx{2(2)}$ generators $T_A$, $A=1,\dots,14$. Only the first set yields a solution and represents the $\SU{2}_R$ symmetry group. The set $\{j_\alpha\}$ is a basis for the Cartan subalgebra of $\mathfrak{su}_2$. The coset representative $\mathbb{L}$ for $\tfrac{\Gx{2(2)}}{\SU{2}\times \SU{2}}$ is given by (7.12.2). The dressed generators $J_\alpha$ are obtained by the action of $\mathbb{L}$ on $j_\alpha$:
\begin{align}
    \label{app_g2_dressed_generators}
    J_\alpha = \mathbb{L} \cdot j_\alpha \cdot \mathbb{L}^{-1},
\end{align}
which we also wrote in terms of the fourteen roots of $\Gx{2(2)}$, detailing how the $\SU{2}_R$ subalgebra is embedded in $\Gx{2(2)}$. The coefficients of $J_\alpha$ give the embedding matrices $m_\alpha^A$:
\begin{align}
    \label{app_g2_embedding_matrices}
    J_\alpha = m_\alpha{}^A T_A.
\end{align}
To extract the $p$-tensor, we choose the generator that corresponds to the $\Uni{1}_R$ inside the $\SU{2}_R$ subalgebra $\{J_\alpha\}$. Finally, the hyper-sector embedding tensor is obtained by 
\begin{align}
    \label{app_g2_ptensor}
    [J_{K_0}, T_A] = {p_{0\,A}}^B T_B,
\end{align}
The $A,B$ indices are lowered and raised with the matrix
\begin{align}
    \label{app_g2_eta}
    \eta_{AB} = \tr(T_A T_B).
\end{align}

\bibliographystyle{elsarticle-num} 
\bibliography{bibfile}






\end{document}